\documentclass[aps,pra,reprint,superscriptaddress]{revtex4-2}
\usepackage{amsmath}
\usepackage{amssymb}
\usepackage{mathtools}
\usepackage{xcolor}
\usepackage{graphicx}
\usepackage[normalem]{ulem}
\usepackage{physics}
\usepackage{braket}

\begin{document}

%Title of paper
\title{Analysis of optical quantum state preparation \\ using photon detectors in the finite-temporal-resolution regime}

% repeat the \author .. \affiliation  etc. as needed
% \email, \thanks, \homepage, \altaffiliation all apply to the current
% author. Explanatory text should go in the []'s, actual e-mail
% address or url should go in the {}'s for \email and \homepage.
% Please use the appropriate macro foreach each type of information

% \affiliation command applies to all authors since the last
% \affiliation command. The \affiliation command should follow the
% other information
% \affiliation can be followed by \email, \homepage, \thanks as well.
\author{Tatsuki Sonoyama}
\email{sonoyama@alice.t.u-tokyo.ac.jp}
\affiliation{Department of Applied Physics, School of Engineering, The University of Tokyo, \\ 7-3-1 Hongo, Bunkyo-ku, Tokyo 113-8656, Japan}
\author{Warit Asavanant}
\affiliation{Department of Applied Physics, School of Engineering, The University of Tokyo, \\ 7-3-1 Hongo, Bunkyo-ku, Tokyo 113-8656, Japan}
\author{Kosuke Fukui}
\affiliation{Department of Applied Physics, School of Engineering, The University of Tokyo, \\ 7-3-1 Hongo, Bunkyo-ku, Tokyo 113-8656, Japan}
\author{Mamoru Endo}
\affiliation{Department of Applied Physics, School of Engineering, The University of Tokyo, \\ 7-3-1 Hongo, Bunkyo-ku, Tokyo 113-8656, Japan}
\author{Jun-ichi Yoshikawa}
\affiliation{Department of Applied Physics, School of Engineering, The University of Tokyo, \\ 7-3-1 Hongo, Bunkyo-ku, Tokyo 113-8656, Japan}
\author{Akira Furusawa}
\email{akiraf@ap.t.u-tokyo.ac.jp}
\affiliation{Department of Applied Physics, School of Engineering, The University of Tokyo, \\ 7-3-1 Hongo, Bunkyo-ku, Tokyo 113-8656, Japan}
\affiliation{Optical Quantum Computing Research Team, RIKEN Center for Quantum Computing, \\ 2-1 Hirosawa, Wako, Saitama 351-0198, Japan}
%\email[]{Your e-mail address}
%\homepage[]{Your web page}
%\thanks{}
%\altaffiliation{}
%\affiliation{}

%Collaboration name if desired (requires use of superscriptaddress
%option in \documentclass). \noaffiliation is required (may also be
%used with the \author command).
%\collaboration can be followed by \email, \homepage, \thanks as well.
%\collaboration{}
%\noaffiliation

\date{\today}

\begin{abstract}
Quantum state preparation is important for quantum information processing. In particular, in optical quantum computing with continuous variables, non-Gaussian states are needed for universal operation and error correction. Optical non-Gaussian states are usually generated by heralding schemes using photon detectors. In previous experiments, the temporal resolution of the photon detectors was sufficiently high relative to the time width of the quantum state, so that the conventional theory of non-Gaussian state preparation treated the detector's temporal resolution as negligible. However, when using various photon detectors including photon-number-resolving detectors, the temporal resolution is non-negligible. In this paper, we extend the conventional theory of quantum state preparation using photon detectors to the finite temporal resolution regime, analyze the cases of single-photon and two-photon preparation as examples, and find that the generated states are characterized by the dimensionless parameter $B$, defined as the product of the temporal resolution of the detectors $\Delta t$ and the bandwidth of the light source $\Delta f$. Based on the results, $B\sim0.1$ is required to keep the purity and fidelity of the generated quantum states high.
\end{abstract}

% insert suggested keywords - APS authors don't need to do this
%\keywords{}

%\maketitle must follow title, authors, abstract, and keywords
\maketitle

% body of paper here - Use proper section commands
% References should be done using the \cite, \ref, and \label commands
\section{Introduction\label{sec:Introduction}}
\begin{figure}[htb]
\includegraphics[width=\columnwidth]{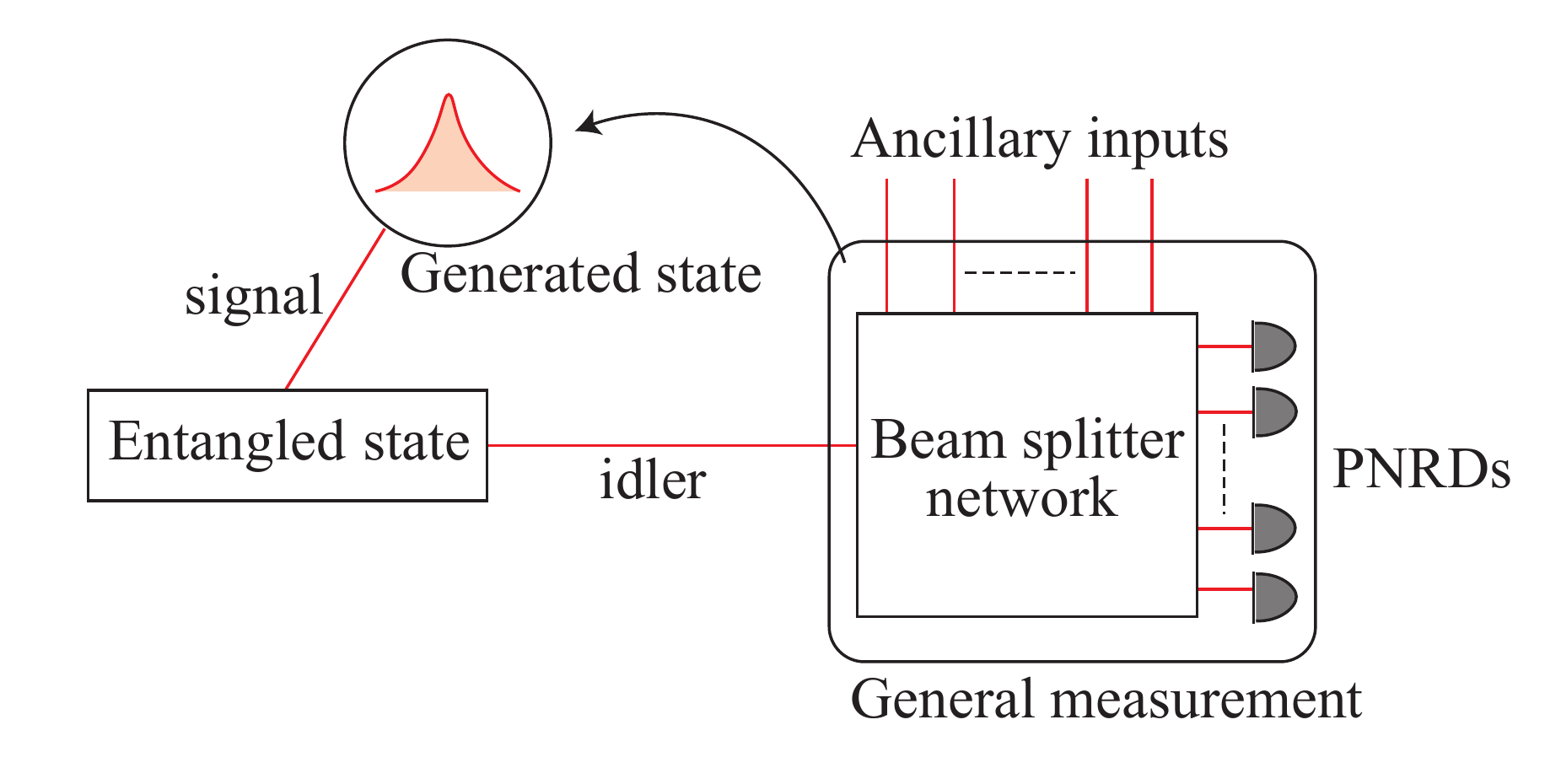}
\caption{Schematic diagram of general non-Gaussian state preparation. When general measurement, which is composed of beam splitter network, ancillary inputs (usually coherent states) and photon-number-resolving detectors (PNRDs), is applied to one mode of the entangled state (idler), non-Gaussian state is generated in the other mode (signal) depending on the measurement result. \label{fig:generalnonGauss}}
\end{figure}

\begin{figure}
\includegraphics[width=\columnwidth]{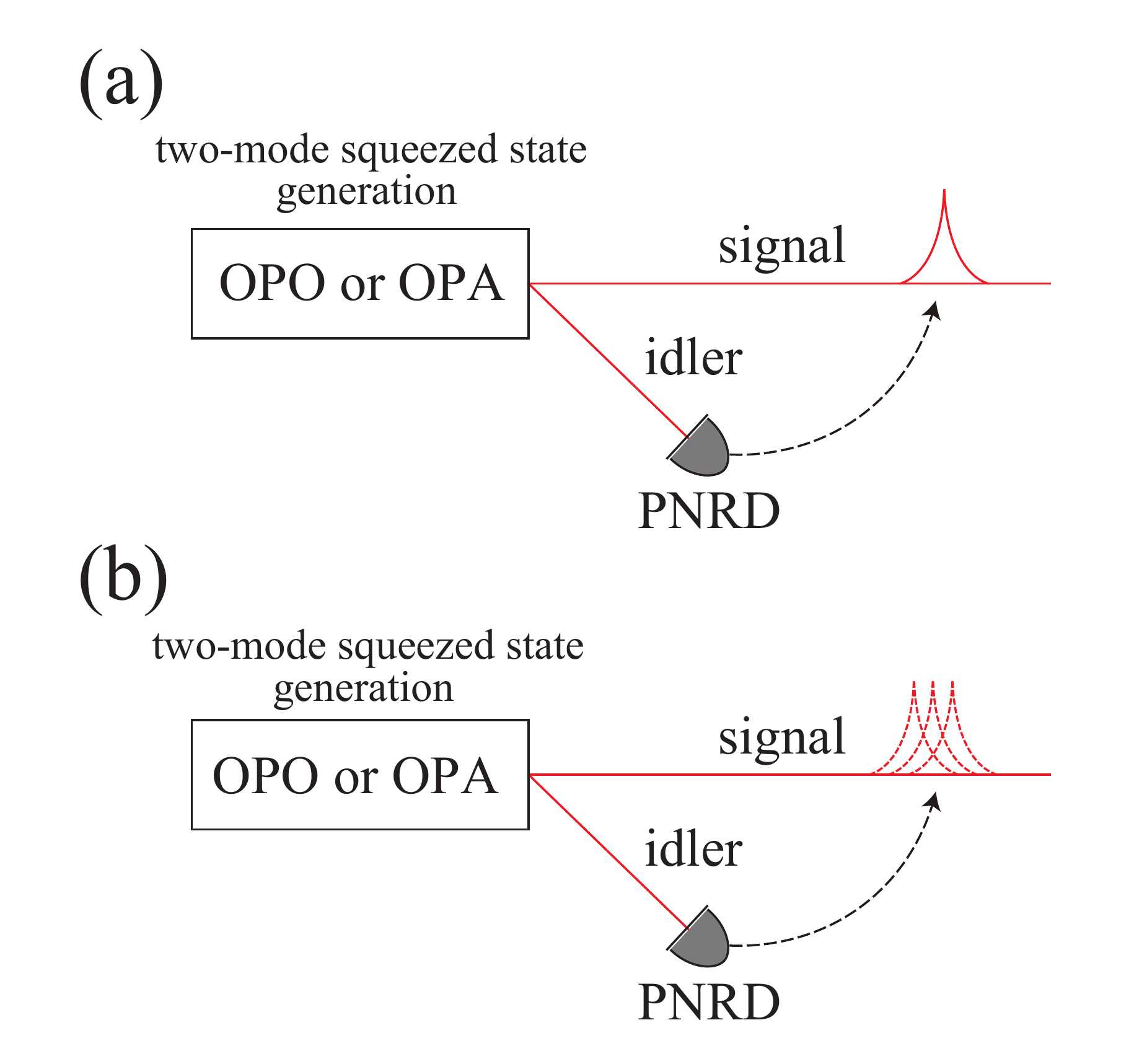}
\caption{Schematic diagram of fock state preparation by a heralding method using a PNRD. Here, the two-mode squeezed state is used as the light source whose bandwidth is $\Delta f$. In this paper, the bandwidth $\Delta f$ is defined by the Half Width at Half Maximum (HWHM). (a)When the temporal resolution of the detector $\Delta t$ is sufficiently high with respect to the bandwidth of the light source $\Delta f$, the generated state can be approximated as a single-mode fock state \cite{Molmer-nonGauss}. (b)When the temporal resolution of the detector $\Delta t$ is not enough with respect to the bandwidth of the light source $\Delta f$, the generated state becomes a multi-mode state in general. \label{fig:statepreparation}}
\end{figure}

In recent years, continuous variable optical quantum computation has been attracting attention due to its scalability. In fact, large-scale computational resources for measurement based quantum computation called cluster states have been generated by a time-domain-multiplexing method using continuous-wave (CW) light sources \cite{Raussendorf-cluster, Nick-cluster,Yokoyama-cluster, Yoshikawa-cluster, Warit-cluster, Mikkel-cluster}, and Gaussian operations using these cluster states have already been implemented \cite{BarameeQC, Larsen-2modeGaussian}. On the other hand, one of the challenges in optical quantum computation is the preparation of non-Gaussian states necessary for error correction and universal operation \cite{Nick-cluster, GKP,GottesmanKnill-theorem,nogo-theorem}. As shown in Fig.\  \ref{fig:generalnonGauss}, non-Gaussian states are generated using an entangled quantum state and photon-number-resolving detectors (PNRDs) \cite{heralding-1,heralding-2}. However, most of previous experiments have generated only simple non-Gaussian states such as a Schr\"{o}dinger's cat state or a fock state using on-off detectors, which discriminate only the presence or absence of photons \cite{Yukawa-fock, Wakui-cat, Warit-cat, Cat-continuous, pulsed-cat}. In order to prepare practical non-Gaussian states for quantum computation such as Gottesman-Kitaev-Preskill (GKP) state \cite{GKP}, it is necessary to use a PNRD capable of multiphoton detection \cite{Fukui-statepreparation,BenchmarkingPNRD}. \\
\indent When using a PNRD, however, the temporal resolution of the detector is an issue. Most of previous experiments for non-Gaussian state preparation used on-off detectors such as Avalanche Photodiodes (APDs) whose timing jitter is several tens of picoseconds \cite{APD}. On the other hand, Transition Edge Sensors (TESs), known as a high-performance PNRD, have a timing jitter of about 4 ns even for a high precision device \cite{4nsTES}. This is a non-negligible value compared to the time width of quantum states in conventional state preparation experiments, which is about several tens of nanoseconds \cite{Yukawa-fock, Wakui-cat, Warit-cat, Cat-continuous}. Although it was shown that the generated state can be regarded as a single-mode pure state defined on a certain temporal mode as shown in Fig.\  \ref{fig:statepreparation}(a) when the temporal resolution is sufficiently high \cite{Molmer-nonGauss}, the generated state when the temporal resolution is non-negligible has not been discussed. \\
\indent In this paper, we extend the theoretical analysis of quantum state preparation using photon detectors to the finite temporal resolution regime. Furthermore, we conduct a specific analysis of single-photon and two-photon states as examples, and show that the generated states are characterized by the dimensionless parameter $B$, which is defined as the product of the detector's temporal resolution $\Delta t$ and the bandwidth of the light source $\Delta f$. Based on the results of this study, the smaller the parameter $B$ is, the higher the purity $P$ and the fidelity $F$ of the generated state become, and $B\sim0.1$ is required for $P, F > 0.9$. \\
\indent The structure of this paper is as follows. First, Sec.\ref{sec:notations} summarizes the notations needed for the theoretical analysis of this paper, and Sec.\ref{sec:idealnonGauss} explains quantum state preparation using a photon detector with sufficiently high temporal resolution. In Sec.\ref{sec:analysis}, we extend the discussion to the case where the temporal resolution of the detector is non-negligible, and analyze the preparation of single-photon state and two-photon state as examples. Then, Sec.\ref{sec:discussion} discusses future quantum state preparation based on the analysis in Sec.\ref{sec:analysis}, and finally, Sec.\ref{sec:conclusion} summarizes this research.

\section{Notations\label{sec:notations}}
Here, we summarize how to describe a quantum state of light beam where longitudinal mode is expressed in time domain \cite{Blowspaper,Takase-PCA}. First, we introduce $\hat{a}(t)$ as an annihilation operator at time $t$, satisfying the commutation relation $[\hat{a}(t),\hat{a}^{\dagger}(t')] = \delta(t-t')$. In the following sections, $\hat{A}(t)$ is also used for convenience centered on the carrier frequency $\omega_0$ of the light source, which is given by
\begin{align}
\hat{A}(t) = \hat{a}(t) e^{i\omega_0 t}.
\end{align}
$\hat{A}(t)$ satisfies the same commutation relation $[\hat{A}(t),\hat{A}^{\dagger}(t')] = \delta(t-t')$ as $\hat{a}(t)$. Also, the creation and annihilation operators $\hat{A}_{f}, \hat{A}^{\dagger}_{f}$ in the temporal mode $f(t)$ are defined using $\hat{A}(t), \hat{A}^{\dagger}(t)$ as
\begin{align}
\hat{A}_{f} = \int_{-\infty}^{\infty} dt f^{*}(t) \hat{A}(t) \,\,\,\, , \,\,\,\, \hat{A}^{\dagger}_{f} = \int_{-\infty}^{\infty} dt f(t) \hat{A}^{\dagger}(t).
\end{align}
By imposing the normalization condition $\int_{-\infty}^{\infty} dt |f(t)|^2 = 1$, the bosonic commutation relation $[\hat{A}_{f},\hat{A}^{\dagger}_{f}]=1$ holds. Using this, for example, the single-photon state $\ket{1_{f}}$ and the two-photon state $\ket{2_{f}}$ in temporal mode $f(t)$ are expressed as follows,
\begin{align}
\label{eq:fock1ft}
\ket{1_{f}} &= \hat{A}_{f}^{\dagger} \ket{\emptyset} = \int_{-\infty}^{\infty} dt f(t) \hat{A}^{\dagger}(t) \ket{\emptyset}, \\
\ket{2_{f}} &= \left(\hat{A}_{f}^{\dagger}\right)^2 \ket{\emptyset} \nonumber \\
&= \iint_{-\infty}^{\infty} dt dt' f(t) f(t') \hat{A}^{\dagger}(t) \hat{A}^{\dagger}(t') \ket{\emptyset}.
\end{align}
Here, the multimode vacuum state $\ket{\emptyset}$ can be defined as a quantum state satisfying the following equation,
\begin{align}
\hat{A}(t) \ket{\emptyset} = 0 \,\,\, , \,\,\, \forall t \in \mathbb{R}.
\end{align}
\indent So far, unbounded and continuous time expression is used, but experimentally finite and discrete time expression is needed. Therefore, we consider dividing finite time width $[-T/2,T/2]$ into $M$ pieces. For this, we define time bin mode function $t_{j}(t)$, which has a finite value only in the $j$-th interval and is expressed as follows,
\begin{align}
t_{j}(t) =
  \begin{cases}
    \displaystyle \sqrt{\frac{M}{T}}\,\,&\frac{(j-1)T}{M}-\frac{T}{2} \leq t \leq \frac{jT}{M}-\frac{T}{2} \\
    \displaystyle 0 &{\rm otherwise},
  \end{cases}
\label{eq:t_j}
\end{align}
where $j$ is an integer from $1$ to $M$. Using this time bin mode function $t_{j}(t)$, the annihilation and creation operators in finite and discrete time $\hat{A}_{t_{j}}$,  $\hat{A}^{\dagger}_{t_{j}}$ are given by
\begin{align}
\hat{A}_{t_{j}} = \int_{-\infty}^{\infty} dt \,t_{j}(t) \hat{A}(t)\,\,,\,\, \hat{A}^{\dagger}_{t_{j}} = \int_{-\infty}^{\infty} dt\, t_{j}(t) \hat{A}^{\dagger}(t).
\label{eq:discreteA}
\end{align}
Considering $\int_{-\infty}^{\infty} dt |t_{j}(t)|^2 = 1$, the bosonic commutation relation $[\hat{A}^{\dagger}_{t_{j}},\hat{A}_{t_j'}]=\delta_{jj'}$ holds. Here, $\delta_{jj'}$ is the Kronecker delta. When the temporal mode function $f(t)$ has negligible values outside the time width $[-T/2,T/2]$ and could be regarded as constant in each interval, $f(t)$ can be expressed as follows, \\
\begin{align}
 f(t) \simeq \sum_{j=1}^{M} f[t_j] t_j(t),
\end{align}
where $f[t_j]  \equiv \sqrt{T/M} f(jT/M - T/2)$. Therefore, the annihilation and creation operators for the temporal mode $f$ can be written as follows,
\begin{align}
 \hat{A}_{f} = \sum_{j=1}^{M} f^{*}[t_j] \hat{A}_{t_j} , \quad \hat{A}^{\dagger}_{f} = \sum_{j=1}^{M} f[t_j] \hat{A}^{\dagger}_{t_j}.
 \label{eq:fdiscreteA}
\end{align}
\indent Finally, we note that the range of integration for the integrals that appear in this paper is from $-\infty$ to $\infty$ unless otherwise specified, and also only one $\int$ in multiple integrals is described.

\section{Quantum state preparation in infinite temporal resolution regime\label{sec:idealnonGauss}}
As a simple example, we consider fock state preparation. In fock state preparation experiment, the quantum state to be measured is a two-mode squeezed state represented as follows,
\begin{align}
\ket{\Psi}_{\rm s,i} = \sqrt{1-q^2} \sum_{n=0}^{\infty} q^n \ket{n}_{\rm s} \ket{n}_{\rm i},
\end{align}
where $q = \tanh r$ is a squeezing parameter. Here, if one mode (idler, i) is measured by a PNRD, the quantum state appearing in the other mode (signal, s) is
\begin{align}
\ket{\psi}_{\rm s} \propto {}_{\rm i} \braket{k|\Psi}_{\rm s,i} \propto \ket{k}_{\rm s} .
\end{align}
This is the simple description of fock state preparation using a PNRD. However, in actual experiments, photon pairs are not generated in two modes, but are generated continuously with time correlation when a CW light source is used. Therefore, the above explanation is not accurate. \\
\indent In the following, we introduce conventional theory of fock state preparation in time domain applicable to the actual experimental situation \cite{Molmer-nonGauss,Yoshikawa-purecat}. As already mentioned, this theoretical analysis treats the temporal resolution of the detectors as infinite. The two-mode squeezed state $
\ket{\Psi} = \hat{S} \ket{\emptyset}_{\rm s} \otimes \ket{\emptyset}_{\rm i}$ generated by parametric down conversion can be expressed by applying the squeezing operator $\hat{S}$ to two-mode vacuum state $\ket{\emptyset}_{\rm i} \otimes \ket{\emptyset}_{\rm s}$. In the case of using a CW pump light, the squeezing operator $\hat{S}$ has time translational symmetry as follows,
\begin{align}
\label{eq:2modesqueezedstate}
& \hat{S} = \exp\left(\int dt dt' r^{*}(t-t') \hat{A}_{\rm s}(t)\hat{A}_{\rm i}(t') \right.  \nonumber \\
& \qquad \qquad \qquad \qquad \qquad  \left.  - r(t-t') \hat{A}_{\rm s}^{\dagger}(t)\hat{A}_{\rm i}^{\dagger}(t')\right).
\end{align}
In the weakly pumped regime, the two-mode squeezed state $\ket{\Psi}_{\rm s,i}$ can be approximated as follows,
\begin{align}
\ket{\Psi}_{\rm s,i} \simeq \left(1 - \int dt dt' r(t-t') \hat{A}_{\rm s}^{\dagger}(t) \hat{A}_{\rm i}^{\dagger}(t') \right) \ket{\emptyset}_{\rm s} \otimes  \ket{\emptyset}_{\rm i}.
\label{eq:1photonapproximation}
\end{align}
When a single photon is detected on the idler mode at $t_0$, the generated state $\ket{\psi_{t_0}}_{\rm s}$ is given by
\begin{align}
\ket{\psi_{t_0}}_{\rm s} \propto {}_{\rm i}\bra{\emptyset}&\hat{A}_{\rm i}(t_0) \hat{S} \ket{\emptyset}_{\rm s} \otimes  \ket{\emptyset}_{\rm i} \simeq \int dt r(t-t_0) \hat{A}_{\rm s}^{\dagger} (t) \ket{\emptyset}_{\rm s}.
\label{eq:fockstate_1}
\end{align}
$\ket{\psi}_{\rm s}$ can be regarded as a single-photon state in the temporal mode function $N[r(t)]$, as in Eq.\  \eqref{eq:fock1ft}. Here, $N[\cdot]$ denotes the normalization of the function. When squeezed light is generated using an Optical Parametric Oscillator (OPO) in the weakly pumped regime, $r(t)$ is determined by the bandwidth of the cavity $\Delta f$, i.e., the bandwidth of the squeezed light, and is a both-side exponential function as follows \cite{Molmer-nonGauss,Yoshikawa-purecat},
\begin{align}
r(t) &= \epsilon \sqrt{2\pi} e^{-\frac{\gamma}{2} |t|} \,\,\, , \,\,\,  \gamma = 4\pi \Delta f.
\label{eq:deltaf}
\end{align}
Here, $\epsilon$ denotes the relative amplitude of the pumped light which is dimensionless. The following analysis in Sec. \ref{sec:fock1} and Sec. \ref{sec:fock2} considers a both-side exponential function as an example. Here we note that this temporal mode function could be modified by introducing assymmerty \cite{Ogawa-assymetricOPO}.

\section{Quantum state preparation in finite temporal resolution regime\label{sec:analysis}}
\subsection{Analysis of general quantum state preparation\label{sec:general}}
In this section, we extend the theory to the case where the detector's temporal resolution is non-negligible with respect to the width of the wave packet. Here we consider a simple case where the idler mode of the two-mode entangled state is detected by a PNRD with finite temporal resolution as shown in Fig.\   \ref{fig:statepreparation}(b). In general, the generated state becomes mixed state when the photon detector has finite temporal resolution. Therefore, we introduce the density operator $\hat{\rho}_{\rm s}$ to describe the generated quantum state. When $n$ photons are detected at time $\{t_k\}_{k=1,2,\cdots,n}$, $\hat{\rho}_{s}$ can be expressed using a Positive Operator-Valued Measure (POVM) operator $\hat{\Pi}_{n,\{t_k\}}$ and the initial two-mode entangled state $\hat{\rho}$ as follows,
\begin{align}
\label{eq:generalrho_s}
\hat{\rho}_{s} = \frac{{\rm tr_i}\left(\hat{\Pi}_{n,\{t_k\}}\hat{\rho}\right)}{{\rm tr}\left(\hat{\Pi}_{n,\{t_k\}}\hat{\rho}\right)},
\end{align}
where ${\rm tr_i}(\cdot)$ denotes partial trace operation for the idler mode (i). Here, the denominator of the right hand side corresponds to the probability density $p(n,\{t_k\})$ for detecting $n$ photons at time $\{t_k\}$, which is expressed as follows,
\begin{align}
p(n,\{t_k\}) = {\rm tr}\left(\hat{\Pi}_{n,\{t_k\}}\hat{\rho}\right).
\end{align}
Also, from the property of the probability density function, the following equation,
\begin{align}
\sum_{n} \int dt_{1} dt_{2} \cdots dt_{n} \hat{\Pi}_{n,\{t_{k}\}} = \hat{I}
\label{eq:POVMnorm}
\end{align}
holds. \\
\indent  Now, we proceed to discuss the specific form of the POVM operator $\hat{\Pi}_{n,\{t_k\}}$. First, we consider $n=1$ as a simple example. By introducing jitter function $g(t)$, the POVM operator $\hat{\Pi}_{1,t_0}$ for detecting single photon at time $t_0$ is expressed as follows,
\begin{align}
\hat{\Pi}_{1,t_{0}} = \int dt g(t-t_0) \hat{A}^{\dagger}(t) \dyad{\emptyset} \hat{A}(t).
\label{eq:Pi1}
\end{align}
The jitter function $g(t)$ is non-negative function determined by the characteristics of the photon detector, satisfying $\int dt \,g(t) = 1$. Also, we can check that when jitter function $g(t)$ is a delta function $\delta(t)$, the above results coincide with those of Sec. \ref{sec:idealnonGauss}. This expression can be easily extended to the multiphoton detection case as follows,
\begin{align}
\hat{\Pi}_{n,\{t_k\}} &= \frac{1}{n!} \int dt'_{1} \cdots dt'_{n} g(t'_{1}-t_1) \cdots g(t'_{n}-t_{n}) \nonumber \\
&\hat{A}^{\dagger}(t'_{1}) \cdots \hat{A}^{\dagger}(t'_{n}) \dyad{\emptyset} \hat{A}(t'_{n}) \cdots \hat{A}(t'_{1}).
\label{eq:generalPOVM}
\end{align}
Note that we are not considering the dead time of the photon detector for simplicity. Also, when the effects of loss and dark counts are considered, the POVM operators are modified such that different photon number terms are mixed. Now, it is possible to calculate the generated quantum state $\hat{\rho}_{\rm s}$ given the specific functional form of $g(t)$ and the quantum state to be measured $\hat{\rho}$.

In general, the generated quantum state $\hat{\rho}_{\rm s}$ becomes multimode state. However, we can determine the temporal mode function of the generated state as the one containing maximal photons. This is the function $f_1(t)$ that maximize $E_{f_1}$ in the following equation,
\begin{align}
 E_{f_1} = {\rm tr}\left(\hat{\rho}_{\rm s}\hat{A}_{f_1}^{\dagger}\hat{A}_{f_1}\right).
\end{align}
 This $f(t)$ can be calculated by diagonalizing the autocorrelation function $E(t,t')$, which is given by
\begin{align}
E(t,t') = {\rm tr}\left(\hat{\rho}_{\rm s}\hat{A}_{\rm s}^{\dagger}(t)\hat{A}_{\rm s}(t')\right).
\label{eq:Econtinuous}
\end{align}
This process is similar to the method used to estimate the temporal mode function by principal component analysis from experimentally obtained data of quadratures \cite{fockstate-PCA,PCA,Takase-PCA}. \\
\indent In the following subsection, the autocorrelation function $E$ is calculated numerically as follows,
\begin{align}
E[i,j] = {\rm tr}\left(\hat{\rho}_{\rm s} \hat{A}^{\dagger}_{t_i} \hat{A}_{t_j} \right) \,\, (1\leq i,j \leq M),
\label{eq:Ediscrete}
\end{align}
where finite and discrete time expression is used as shown in Eq.\  \eqref{eq:t_j}--\eqref{eq:fdiscreteA}. As for the jitter function $g(t)$, we analyze the cases where $g(t)$ are Gaussian functions or rectangular functions with various jitter width $\Delta t$. \\

\subsection{Single-photon state preparation\label{sec:fock1}}
\begin{figure}
 \begin{center}
 \includegraphics[width=\columnwidth]{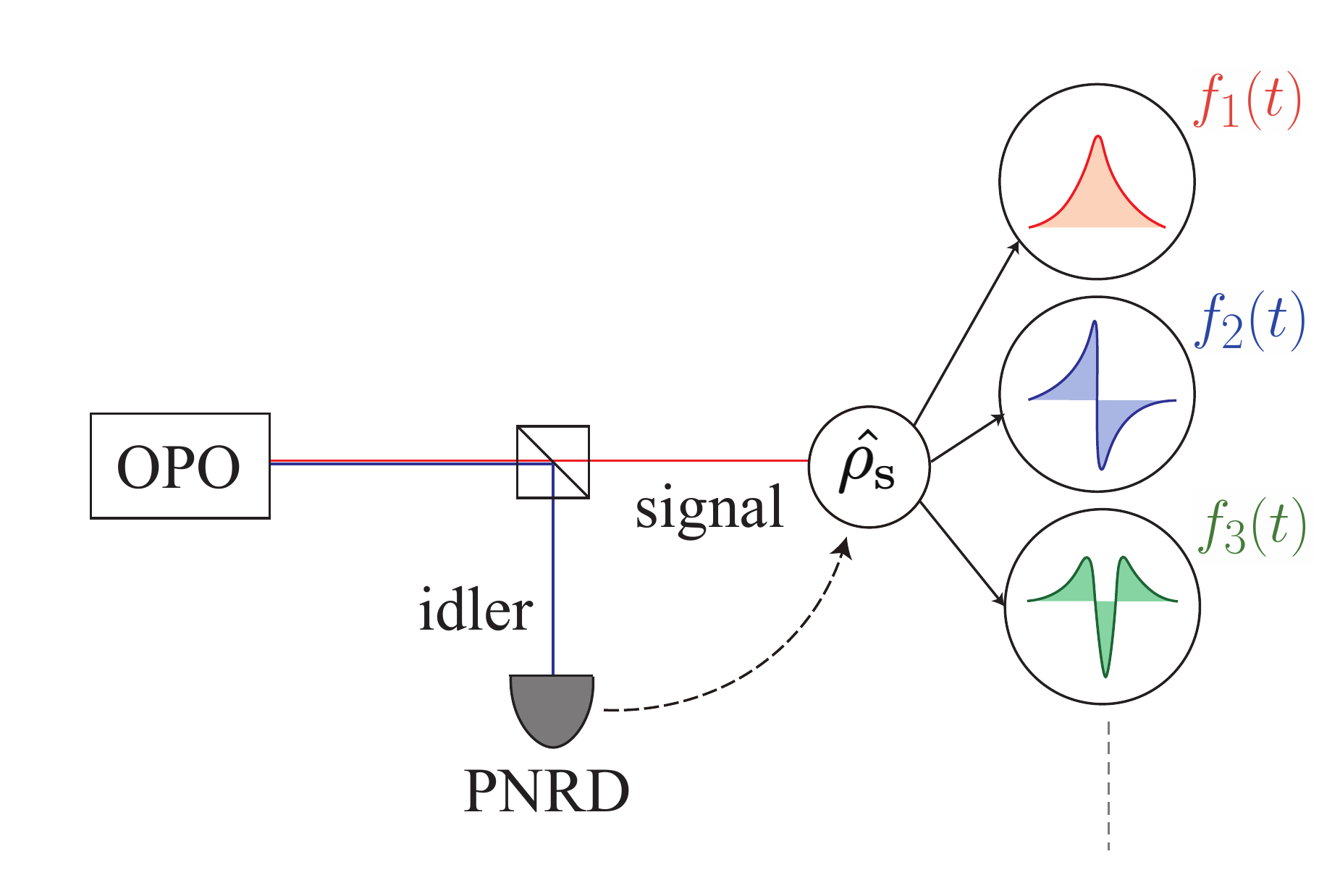}
 \end{center}
 \caption{Schematic diagram of the single-photon state generated when the detector's temporal resolution is non-negligible. This figure shows the generated state is a multimode state and it can be decomposed to single-mode single-photon states $\ket{\psi_{k}}$ whose temporal mode function is $f_{k}(t)$.  \label{fig:multimodestate}}
\end{figure}

Here, we consider single-photon state preparation. The quantum state to be measured $\hat{\rho}$ is a two-mode squeezed state generated by a CW light source in the weakly pumped regime, and is expressed as $\hat{\rho} = \dyad{\Psi}$ using $\ket{\Psi}$ in Eq.\  \eqref{eq:1photonapproximation}. The generated quantum state $\hat{\rho}_{s}$ are calculated using this quantum state $\hat{\rho}$ and the POVM operator $\hat{\Pi}_{1,t_c}$ in Eq.\ \eqref{eq:Pi1} corresponding to one photon detection at time $t_c$ as follows,
\begin{align}
\hat{\rho_{\rm s}} \propto {\rm tr_{i}}\left(\hat{\Pi}_{1,t_c} \hat{\rho}\right) = {\rm tr_{i}}\left(\hat{\Pi}_{1,t_c} \dyad{\Psi} \right).
\label{eq:multimodestate_calc}
\end{align}
In this paper, two cases are considered for the jitter function $g(t)$: one is a rectangular function $g_{r}(t)$ that is uniform within $\Delta t$ and the other is a Gaussian function $g_{g}(t)$ whose Full Width at Half Maximum (FWHM) matches $\Delta t$, as follows,
\begin{align}
  g_{r}(t) \propto
  \begin{cases}
    \displaystyle 1\,\,&|t|\leq \frac{\Delta t}{2}  \\
    \displaystyle 0 &{\rm otherwise}
  \end{cases}
  \,\,\, &{\rm or} \,\,\, g_{g}(t) \propto \exp(-\frac{4\ln 2}{(\Delta t)^2} t^2).
  \label{eq:singlephotongt}
\end{align}
From the above, the generated quantum state $\hat{\rho}_{\rm s}$ can be calculated as
\begin{align}
\hat{\rho}_{\rm s} &\propto {\rm tr_{i}}\left(\hat{\Pi}_{1,t_c} \hat{\rho}\right) \nonumber \\
&= {\rm tr_{i}}\left(\int dt g(t-t_c)\hat{A}_{\rm i}^{\dagger}(t)\dyad{\emptyset}_{\rm i}\hat{A}_{\rm i}(t)\dyad{\Psi}\right) \nonumber \\
&= \int dt dt' dt''  g(t-t_c) r(t'-t)     \nonumber \\
&\qquad \qquad \quad \quad r^{*}(t''-t) \hat{A}^{\dagger}_{\rm s}(t') \dyad{\emptyset}_{\rm s} \hat{A}_{\rm s}(t'')\,\, \nonumber \\
&= \int dt g(t-t_c) \dyad{\psi_{t}}_{\rm s},
\label{eq:multimodestate_calc}
\end{align}
where $\ket{\psi_{t}}$ shown in Eq.\ \eqref{eq:fockstate_1} is used. The generated state is in the form of $\dyad{\psi_{t}}$ integrated over time $t$, which is a classical mixed state. In the case of a single-photon state considering here, instead of diagonalizing the autocorrelation matrix $E(t,t')$ in Eq.\ \eqref{eq:Econtinuous}, the density matrix $\hat{\rho_{\rm s}}$ can be directly diagonalized as shown in Fig.\ \ref{fig:multimodestate} and in the following equation,
\begin{align}
\hat{\rho_{\rm s}} = \sum_{k} \lambda_{k} \dyad{\psi_{k}},
\label{eq:diagonalization1}
\end{align}
where $\ket{\psi_k}$ satisfies $\Braket{\psi_{k}|\psi_{k'}} = \delta_{kk'}$. Here, the single-photon states $\ket{\psi_k}$ are defined as $\ket{\psi_{k}} = \int dt f_k(t) \hat{A}^{\dagger}(t) \ket{\emptyset}$, where $f_k(t)$ are mutually orthogonal eigenmode functions.
The first eigenmode function $f_1(t)$ corresponding to the biggest eigenvalue $\lambda_1$ can be regarded as the optimal temporal mode function of the generated quantum state. Also, fidelity $F$ between the generated state and the target state $\ket{\psi_1}$ and purity $P$ of the generated state can be calculated from the eigenvalues as follows,
\begin{align}
 F = \bra{\psi_1}\hat{\rho}_{\rm s}\ket{\psi_1} = \lambda_1 \, , \,  P = {\rm tr}\left(\hat{\rho}^2\right) = \sum_{k} \lambda_{k}^2.
\end{align}

\begin{figure}
\includegraphics[width=\columnwidth]{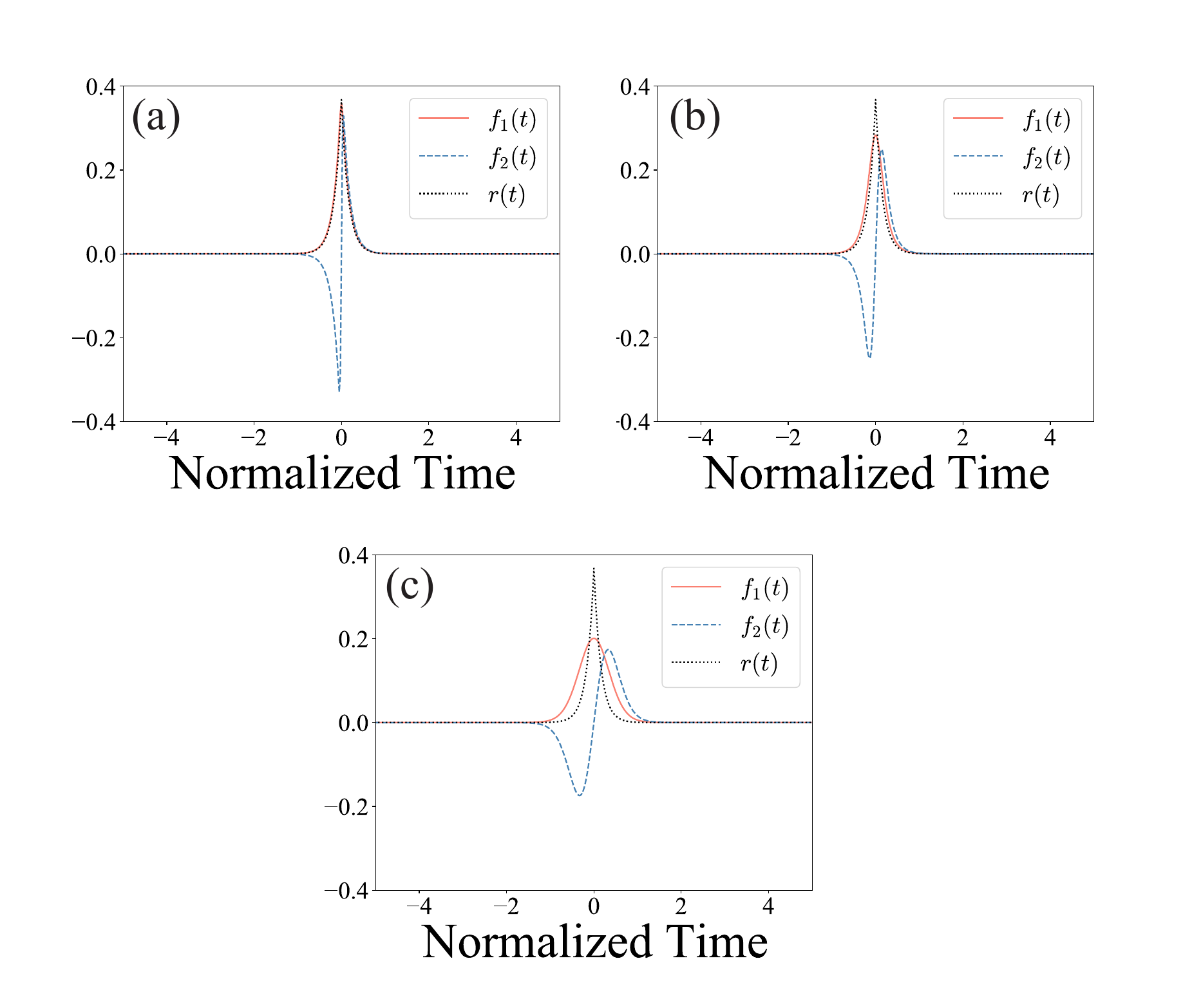}%
\caption{Numerically calculated temporal mode functions of a single-photon state. In this calculation, the quantum state $\hat{\rho}_{\rm s}$ is described in finite and discrete time. The time width $T$ is set to $10$, which is normalized by the bandwidth of the light source and the number of divisions $M$ is set to $800$. In this figure, only the first and second eigenmodes $f_1(t),f_2(t)$ are shown when $g(t)$ are Gaussian functions. For clarity, $r(t)$ is also plotted, appropriately multiplied by a constant. (a)The case of $B=0.05$. (b)The case of $B=0.25$. (c)The case of $B=1.0$. Here, we note that the temporal mode functions only depend on the dimensionless parameter $B=\Delta f \times \Delta t$ because the time is normalized by the bandwidth of the light source $\Delta f$.\label{fig:timemodefunction}}
\end{figure}
In the following, the numerical analysis of the temporal mode function, fidelity and purity are shown depending on a dimensionless parameter $B$. Here, $B = \Delta f \times \Delta t$ is defined as the product of the light source bandwidth $\Delta f$ in Eq.\ \eqref{eq:deltaf} and the detector's temporal resolution $\Delta t$ in Eq.\ \eqref{eq:singlephotongt}.

Fig.\  \ref{fig:timemodefunction} shows the numerical calculation of the temporal mode functions when $g(t)$ are Gaussian functions with $B=0.05,0.25,1.0$. Here, the photon detection time $t_c$ is set to 0 for simplicity. When $B$ is small, the first temporal mode function almost matches the both-side exponential function $r(t)$, but as $B$ becomes larger, the first temporal mode becomes like a blunted both-side exponential function. The similar results are obtained for the case where $g(t)$ are rectangular functions. \\
\indent Next, the purity and the fidelity of the generated state are shown in Fig.\  \ref{fig:BP}(a) and Fig.\  \ref{fig:BF}(a). These plots show that the smaller the value of $B$ is, the closer the fidelity and purity are to the ideal value 1. The plots also show that, for example, if we want to keep the purity and fidelity of the generated state above 0.9, we need $B < 0.1$ when $g(t)$ are Gaussian functions. For further detailed calculations, see Appendix A.

\subsection{Two-photon state preparation\label{sec:fock2}}
\begin{figure}
\includegraphics[width=\columnwidth]{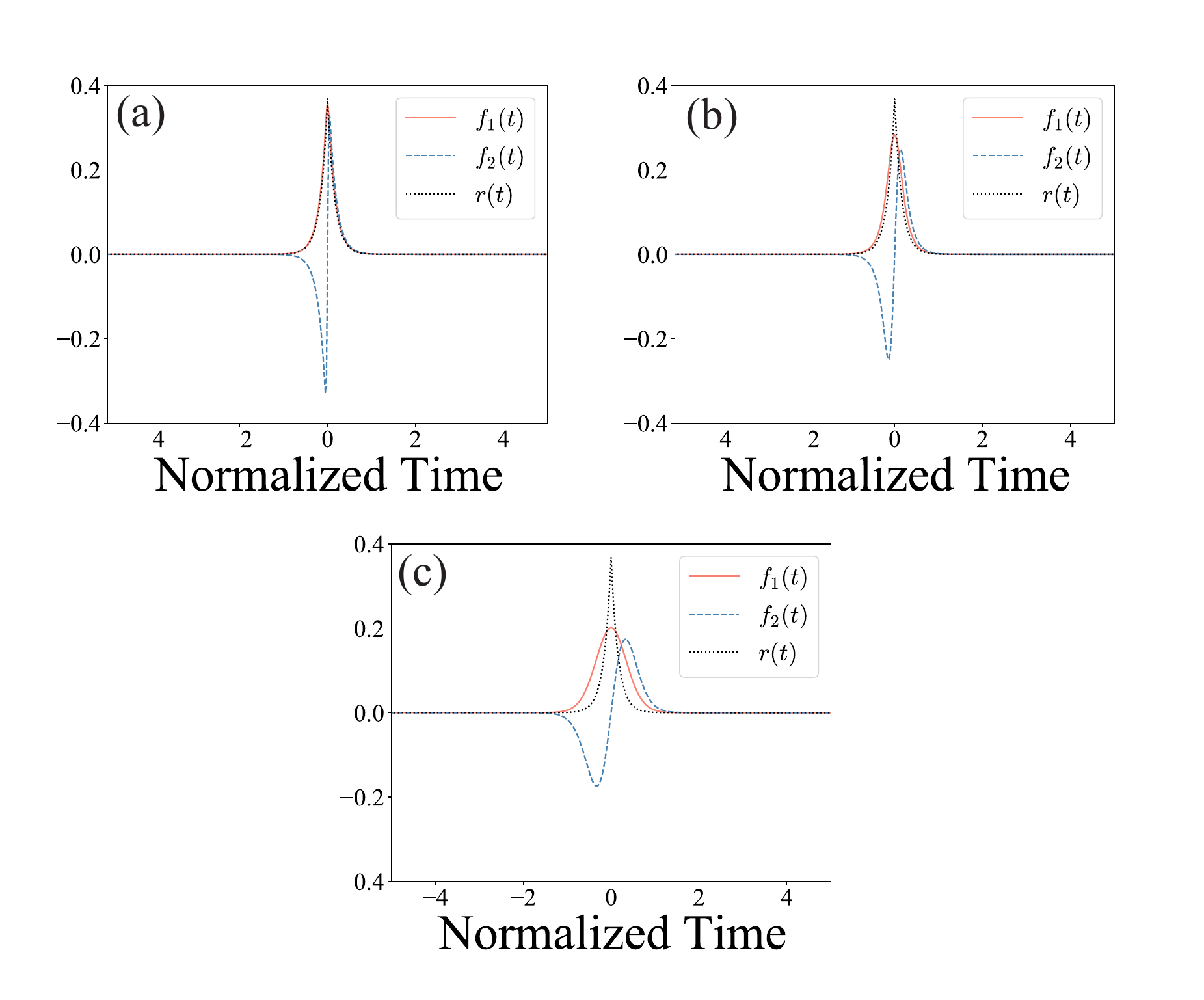}
\caption{Numerically calculated temporal mode functions of a two-photon state. In this calculation, the autocorrelation function $E(t,t')$ is described in finite and discrete time. The time width $T$ is set to $10$, which is normalized by the bandwidth of the light source and the number of divisions $M$ is set to $800$. In this figure, only the first and second eigenmodes $f_1(t),f_2(t)$ are shown when $g(t)$ are Gaussian functions. For clarity, $r(t)$ is also plotted, appropriately multiplied by a constant. (a)The case of $B=0.05$. (b)The case of $B=0.25$. (c)The case of $B=1.0$. Here, we note that the temporal mode functions only depend on the dimensionless parameter $B=\Delta f \times \Delta t$ because the time is normalized by the bandwidth of the light source $\Delta f$.}
\label{fig:timemodefunction2}
\end{figure}

Next, let us consider two-photon state preparation as an example of a multiphoton state using the POVM operator $\hat{\Pi}_{2,{\{t_1,t_2\}}}$. Here we only analyze the case of $t_1 = t_2 = t_c$ in order to consider only the effects of timing jitter. The generated quantum state $\hat{\rho}_{\rm s}$ is given by
\begin{align}
\hat{\rho_{\rm s}} \propto {\rm tr_{i}}\left(\hat{\Pi}_{2,t_c} \hat{\rho}\right) = {\rm tr_{i}}\left(\hat{\Pi}_{2,t_c} \dyad{\Psi} \right).
\label{eq:POVM2photon}
\end{align}
In order to analyze the case of two-photon state preparation, we extend the initial entangled state $\ket{\Psi}$ up to four photon terms as follows,
\begin{align}
\begin{split}
\ket{\Psi} &\simeq \left(1 - \int dt dt' r(t-t') \hat{A}_{\rm s}^{\dagger}(t)\hat{A}_{\rm i}^{\dagger}(t') \right.\\
&\quad \left.+\frac{1}{2} \int dt dt' dt'' dt''' r(t-t') r(t''-t''')\right.\\
&\qquad \qquad \qquad \left. \hat{A}_{s}^{\dagger}(t) \hat{A}_{i}^{\dagger}(t') \hat{A}_{s}^{\dagger}(t'') \hat{A}_{i}^{\dagger}(t''') \right) \ket{\emptyset}_{\rm s} \otimes \ket{\emptyset}_{\rm i}.
\end{split}
\label{eq:1photonapproximation-2}
\end{align}
Next, the POVM operator $\hat{\Pi}_{2,t_c}$ can be expressed as
\begin{align}
  \hat{\Pi}_{2,t_c} &= \frac{1}{2} \int dt dt' g(t-t_c) g(t'-t_c) \nonumber \\
  &\qquad \hat{A}_{\rm i}^{\dagger}(t) \hat{A}_{\rm i}^{\dagger}(t') \dyad{\emptyset}_{\rm i}\hat{A}_{\rm i}(t') \hat{A}_{\rm i}(t),
\end{align}
where $g(t)$ is $g_{\rm g}(t)$ or $g_{\rm r}(t)$ defined in Eq.\  \eqref{eq:singlephotongt}. Based on the above, the generated quantum state $\hat{\rho}_{\rm s}$ can be calculated as
\begin{align}
\hat{\rho_{\rm s}} \propto &\int dt_1 dt_2 g(t_1-t_c) g(t_2-t_c) \int dt dt' dt'' dt''' \nonumber \\
&\, r(t-t_1) r(t'-t_2) r^{*}(t''-t_1) r^{*}(t'''-t_2) \nonumber \\
&\quad \hat{A}_{\rm s}^{\dagger}(t) \hat{A}_{\rm s}^{\dagger}(t') \dyad{\emptyset}_{\rm s} \hat{A}_{\rm s}(t'') \hat{A}_{\rm s}(t''').
\end{align}
In the case of a multiphoton state, the diagonalization of density matrix doesn't correspond to the orthogonalization of the temporal mode functions. Therefore, the temporal mode functions are obtained by the diagonalization of the autocorrelation matrix $E$ as described in Sec.\ref{sec:general}. Fig.\  \ref{fig:timemodefunction2} shows the temporal mode functions when $g(t)$ are Gaussian functions. As in the case of the single-photon state, the first temporal mode becomes like a blunted both-side exponential function as $B$ increases. The similar results are obtained for the case where $g(t)$ are rectangular functions.
Also, the fidelity and the purity of the generated quantum state $\rho_{\rm s}$ are calculated as shown in Fig.\  \ref{fig:BP}(b) and Fig.\  \ref{fig:BF}(b). Here, Fig.\  \ref{fig:BP} and Fig.\  \ref{fig:BF} show that the purity and the fidelity are determined only by $B$, and the smaller the value of $B$ is, the closer the value of $P$ and $F$ is to 1. For example, if $g(t)$ are Gaussian functions, $B<0.15$ is required to satisfy $P > 0.9$ and $F > 0.9$. Surprisingly, we can see the requirement for $B$ does not become stronger when photon number is increased. For further detailed calculations, see Appendix B.

\begin{figure}
 \begin{center}
 \includegraphics[width=\columnwidth]{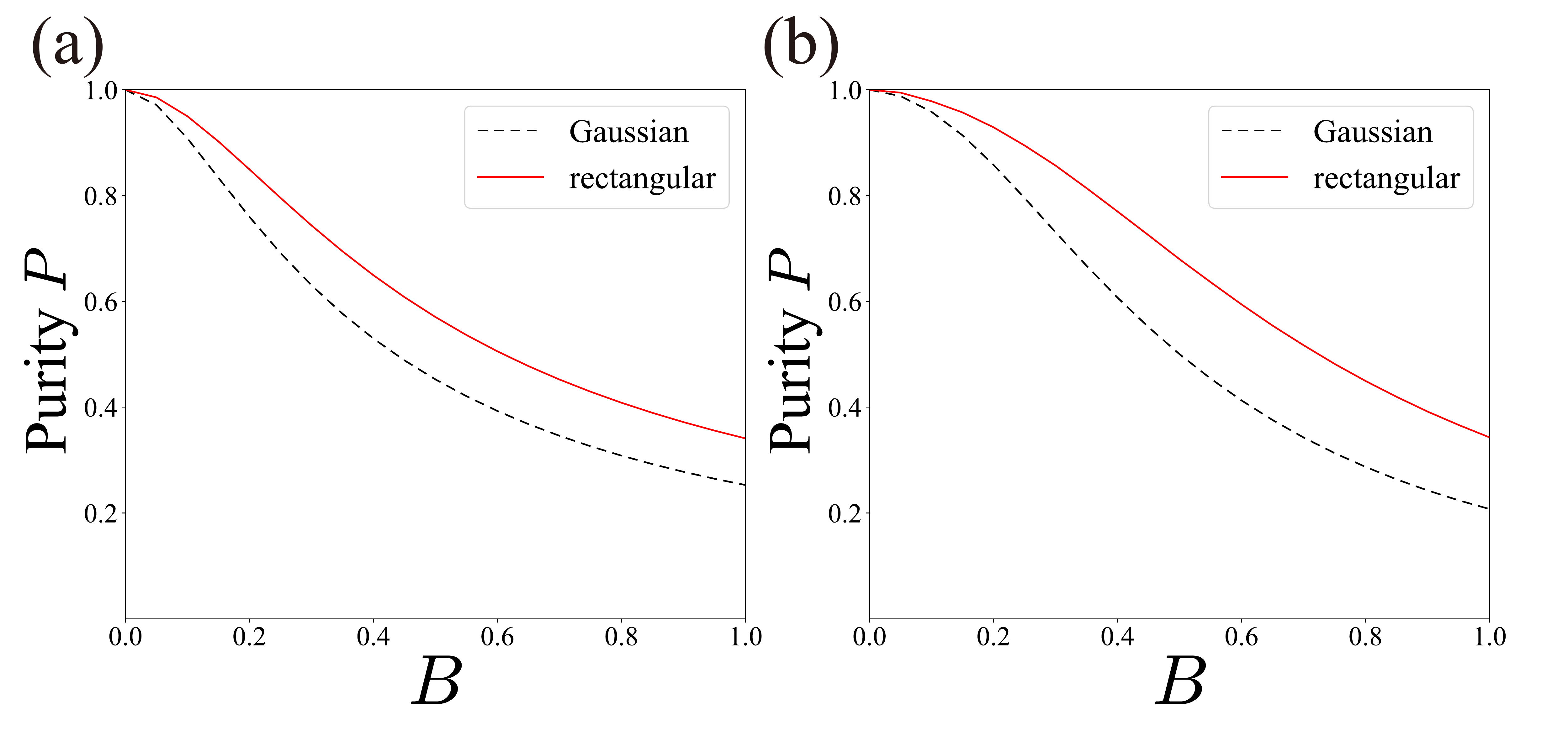}
 \end{center}
 \caption{The plot of the purity of the generated state against $B=\Delta f \times \Delta t$ where $g(t)$ are Gaussian functions and rectangular functions. (a)The purity of single-photon state. the number of divisions $M$ and the time width $T$ are set to $M=800$ and $T=10$. (b) The purity of two-photon state. Due to the speed of numerical calculations, the number of divisions $M$ and the time width $T$ are set to $M=100$ and $T=4$. \label{fig:BP}}
 \label{fig:BP}
\end{figure}

\begin{figure}
 \begin{center}
 \includegraphics[width=\columnwidth]{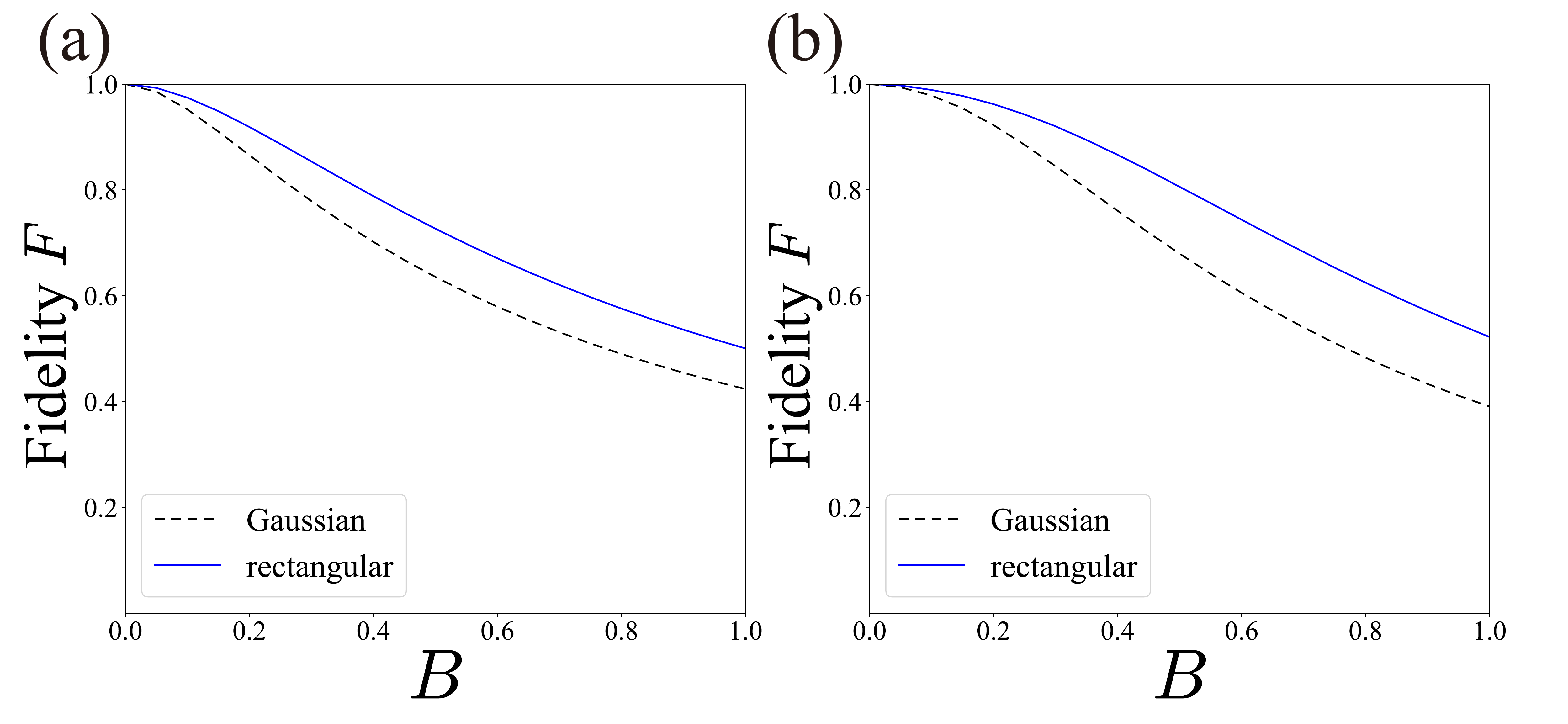}
 \end{center}
 \caption{The plot of the fidelity of the generated state against $B=\Delta f \times \Delta t$ where $g(t)$ are Gaussian functions and a rectangular functions. (a)The fidelity of single-photon state. the number of divisions $M$ and the time width $T$ are set to $M=800$ and $T=10$. (b) The fidelity of two-photon state. Due to the speed of numerical calculations, the number of divisions $M$ and the time width $T$ are set to $M=100$ and $T=4$. \label{fig:BF}}
 \label{fig:BF}
\end{figure}

\section{Discussion\label{sec:discussion}}
The above analysis showed that $B \lesssim 0.1$ should be satisfied to keep the fidelity and the purity of the quantum state. This means that the upper limit of the available light source bandwidth $\Delta f$ is about $1/(10\Delta t)$ for the temporal resolution $\Delta t$ of the detector. Here, the bandwidth $\Delta f$ of the light source is an important quantity that determines the quantum computation speed \cite{TakedaFurusawa}. In measurement-based optical quantum computation, the computation speed is determined by the time required for measurement, that is, the time width of the wave packet containing quantum information. Since the time width of the wave packet is about the reciprocal of the light source bandwidth $\Delta f $, the upper limit of the clock frequency is determined by the light source bandwidth $\Delta f$. Based on the above discussion, it can be also said that the clock frequency of quantum computation is limited by about $1/(10\Delta t)$. Therefore, the temporal resolution of the detector $\Delta t$ needs to be small in order to realize faster quantum computation in the future, and the required value of the detector's temporal resolution can be obtained using the analysis in Sec.\ref{sec:analysis}. For example, if we want to perform quantum computation with a clock frequency of $1$ GHz, we need to use a detector with a temporal resolution of less than $100$ ps. Of course, this upper limit depends on the quantum state used and the required value of the fidelity or purity. Thus, we expect that future works based on this paper will reveal a relationship between $B$ and the fidelity $F$ or purity $P$ for fault-tolerant quantum states such as GKP qubits. \\
\indent Now, we discuss current photon-number-resolving devices for optical quantum computation. First, Transition Edge Sensors (TESs), which have already been introduced, are capable of detecting up to about 20 photons and its temporal resolution is about 4 ns for a high precision one \cite{4nsTES}. While TESs can resolve a large number of photons, its temporal resolution is an issue. Recently, it was also shown that Superconducting Nanostrip Photon Detectors (SNSPDs) can be used as PNRDs \cite{SNSPDPNRD-first,SNSPDPNRD}. Although it can only detect up to 5 photons at present, it has a high temporal resolution of less than 100 ps \cite{SNSPDPNRD}. In this way, photon-number-resolving devices are still under research and development, and further performance improvements are expected in the future.

\section{Conclusion\label{sec:conclusion}}
In this work, we propose a general analysis method for quantum state preparation using a photon detector with finite temporal resolution. Also, in the analysis of single-photon state and two-photon state preparation, the temporal mode function, fidelity and purity are shown to be dependent on the bandwidth of the light source $\Delta f$ and the temporal resolution of the detector $\Delta t$. As a result, it is shown that the smaller $B(=\Delta f\times \Delta t)$ is, the higher the purity and fidelity of the generated state become. The proposed analysis method is important for non-Gaussian state preparation because it shows required temporal resolution of PNRD to keep the purity and fidelity of the generated quantum state. \\

\appendix{}
\section{Detailed calculations in the analysis of single-photon state preparation\label{sec:AppendixA}}
In sec.\ref{sec:fock1}, the eigenmode functions and eigenvalues were obtained by diagonalizing the generated quantum state, and the detailed calculations are shown here. First, Eq.\  \eqref{eq:multimodestate_calc}, which indicates the generated state $\hat{\rho}_{\rm s}$, can be further organized as follows,
\begin{align}
\hat{\rho}_{\rm s} &= \int dt' dt'' \rho(t',t'') \hat{A}^{\dagger}_{\rm s}(t') \dyad{\emptyset}_{\rm s} \hat{A}_{\rm s}(t''), \nonumber \\
\rho(t',t'') & = \int dt g(t) r(t'-t) r^{*}(t''-t)\,\,,
\end{align}
where the photon detection time $t_c = 0$ is assumed. This $\rho(t',t'')$ can be calculated analytically as follows when $g(t)$ is both a rectangular function and a Gaussian function.  Here, we label $\rho(t',t'')$ as $\rho_{\rm g}(t',t'')$ when $g(t)$ is Gaussian, and $\rho_{\rm r}(t',t'')$ when $g(t)$ is rectangular.
\begin{widetext}
\begin{itemize}
\item When $g(t)$ is a Gaussian function as shown in Eq.\  \eqref{eq:singlephotongt} $\left(g(t) \propto \exp(-\mu t^2), \mu = \frac{4\ln 2}{(\Delta t)^2}\right)$, $\rho_{\rm g}(t',t'')$ is given by\\
\begin{align}
\rho_{\rm g}(t',t'') &\propto \exp\left(-\frac{\gamma}{2}\left(t'+t'' \right) + \frac{\gamma^2}{4\mu}\right) \left(1+{\rm erf}\left(\sqrt{\mu}(t''-\frac{\gamma}{2\mu})\right)\right) +\exp\left(\frac{\gamma}{2}\left(t''-t' \right)\right) \left({\rm erf}\left(\sqrt{\mu}t'\right)-{\rm erf}\left(\sqrt{\mu}t''\right)\right) \nonumber \\
& \qquad \qquad \qquad \qquad \qquad \qquad \qquad +\exp\left(\frac{\gamma}{2}\left(t'+t'' \right)+\frac{\gamma^2}{4\mu}\right) \left(1-{\rm erf}\left(\sqrt{\mu}(t'+\frac{\gamma}{2\mu})\right)\right) \quad (t'>t''),
\end{align}
where normalization constant is ignored. Here, since $\rho_{\rm g}(t',t'')$ is symmetric with respect to the interchange of variables, we consider only the case of $t'>t''$. Also, ${\rm erf}(x)$ is the error function which is defined as
\begin{align}
{\rm erf}(x) = \frac{2}{\sqrt{\pi}}\int_{0}^{x} e^{-t^2} dt.
\end{align}
\item When $g(t)$ is a rectangular function as shown in Eq.\  \eqref{eq:singlephotongt}, $\rho_{\rm r}(t',t'')$ is given by\\
\begin{align}
  \rho_{\rm r}(t',t'') \propto
  \begin{cases}
    \displaystyle &\exp(-\frac{\gamma}{2} (t'+t''-\Delta t)) - \exp(-\frac{\gamma}{2} (t'+t''+\Delta t)) \qquad t'>\frac{\Delta t}{2}, t''>\frac{\Delta t}{2}  \\
    \displaystyle &\exp(-\frac{\gamma}{2} (t'-t'')) - \exp(-\frac{\gamma}{2} (t'+t''+\Delta t)) + \gamma(\frac{\Delta t}{2} - t'') \exp(-\frac{\gamma}{2} (t'-t'')) \qquad t'>\frac{\Delta t}{2}, -\frac{\Delta t}{2}<t''<\frac{\Delta t}{2} \\
    \displaystyle &\gamma \Delta t \exp(-\frac{\gamma}{2}(t'-t''))\qquad t'>\frac{\Delta t}{2}, t''<-\frac{\Delta t}{2} \\
    \displaystyle &2\exp(-\frac{\gamma}{2} (t'-t'')) - \exp(-\frac{\gamma}{2} (t'+t''+\Delta t)) - \exp(\frac{\gamma}{2} (t'+t''-\Delta t)) + \gamma (t'-t'')  \exp(-\frac{\gamma}{2}(t'-t''))   \\
    \displaystyle &\qquad \qquad \qquad \qquad \qquad \qquad \qquad \qquad \qquad \qquad \qquad \qquad \qquad \qquad -\frac{\Delta t}{2}<t'<\frac{\Delta t}{2}, -\frac{\Delta t}{2}<t''<\frac{\Delta t}{2} \\
    \displaystyle &\exp(\frac{\gamma}{2} (-t'+t'')) - \exp(\frac{\gamma}{2} (t'+t''-\Delta t)) + \gamma (t'+\frac{\Delta t}{2}) \exp(-\frac{\gamma}{2} (t'-t'')) \\
     \displaystyle &\qquad \qquad \qquad \qquad \qquad \qquad \qquad \qquad \qquad \qquad \qquad \qquad \qquad \qquad -\frac{\Delta t}{2}<t'<\frac{\Delta t}{2}, t'' < -\frac{\Delta t}{2}\\
    \displaystyle &\exp(\frac{\gamma}{2} (t'+t''+\Delta t)) - \exp(\frac{\gamma}{2} (t'+t''-\Delta t)) \qquad t' <  -\frac{\Delta t}{2}, t'' < -\frac{\Delta t}{2} ,
  \end{cases}
\end{align}
where normalization constant is ignored. As when $g(t)$ is a Gaussian function, $\rho_{\rm r}(t',t'')$ is symmetric with respect to the interchange of variables, so here we consider only the case of $t'>t''$.
\end{itemize}
\end{widetext}
When $\rho(t',t'')$ is diagonalized, the generated state $\hat{\rho}_{\rm s}$ is expressed as a mixture of single-photon states in temporal mode functions $f_{k}(t)$, as shown in Eq.\  \eqref{eq:diagonalization1}. In Sec.\ref{sec:fock1}, $\rho(t',t'')$ is discretized in a finite time span for simplicity and diagonalization is performed numerically. Some of these results are shown in Fig.\  \ref{fig:timemodefunction}, Fig.\  \ref{fig:BP}(a) , and Fig.\  \ref{fig:BF}(a).

\section{Detailed calculations in the analysis of two-photon state preparation\label{sec:AppendixB}}
In Sec.\ref{sec:fock2}, we obtained the autocorrelation function $E(t,t')$ of the generated two-photon state and diagonalized it to obtain the temporal mode function, and here we show the detailed calculation. As in the case of the single-photon state, the generated state $\hat{\rho}_{s}$ can be further organized as
\begin{align}
& \hat{\rho}_{\rm s} = \int dt dt' dt'' dt''' \rho(t',t'',t''',t'''') \nonumber \\
&\qquad \qquad \qquad \hat{A}^{\dagger}_{\rm s}(t) \hat{A}^{\dagger}_{\rm s}(t') \dyad{\emptyset}_{\rm s} \hat{A}_{\rm s}(t'') \hat{A}_{\rm s}(t'''), \nonumber \\
& \rho_{k}(t,t',t'',t''') = \int dt_1 dt_2 g(t_1)g(t_2) \nonumber \\
& \qquad r(t-t_1) r(t'-t_2) r^{*}(t''-t_1) r^{*}(t'''-t_2)\,\,,
\end{align}
where the photon detection time $t_c = 0$ is assumed. This $\rho(t,t',t'',t'')$ can be expressed as follows when $g(t)$ is both a rectangular function and a Gaussian function using $\rho_{\rm g}$ and $\rho_{\rm r}$ calculated in Appendix.\ref{sec:AppendixA}.
\begin{align}
\rho_{k} (t,t',t'',t'') = \rho_{k} (t,t'') \rho_{k} (t',t''') \qquad (k=g,r).
\end{align}
Here, we treat the density matrix $\hat{\rho}_{\rm s}$ by discretizing the interval [$-T/2, T/2$] into $M$ pieces so that it can be handled numerically. Using the creation operator in the finite and discrete time $\hat{A}_{t_j}^{\dagger}$ defined in Eq.\  \eqref{eq:discreteA}, the density matrix $\hat{\rho}_{\rm s}$ can be approximated as follows.
\begin{gather}
\hat{\rho_{\rm s}} \simeq \sum_{a,b,c,d=1}^{M} \frac{T^2}{M^2} \rho[a,b,c,d] \hat{A}^{\dagger}_{t_a} \hat{A}^{\dagger}_{t_b} \dyad{\emptyset} \hat{A}_{t_c} \hat{A}_{t_d},
\end{gather}
where the density matrix elements are considered to have negligible values beyond the time width [$-T/2,T/2$] and do not vary significantly within a single interval. Here, density matrix element $\rho[a,b,c,d]$ is given by
\begin{gather}
\rho[a,b,c,d] = \rho\left(\frac{aT}{M}-\frac{T}{2} ,\frac{bT}{M}-\frac{T}{2} ,\frac{cT}{M}-\frac{T}{2} ,\frac{dT}{M}-\frac{T}{2}\right)\nonumber \\
(1 \leq a,b,c,d \leq M).
\end{gather}
Using this discretized density matrix, the autocorrelation matrix $E[i,j]$ can be calculated as
\begin{align}
&E[i,j] \propto \sum_{t=1}^{M} \left(\rho[i,t,j,t] + \rho[i,t,t,j] \right. \nonumber \\
&\qquad \qquad \left. + \rho[t,i,j,t] + \rho[t,i,t,j] \right)\,\, (1 \leq i,j \leq M).
\label{eq:E}
\end{align}
In Sec.\ref{sec:fock2}, the temporal mode functions are obtained by numerically calculating $E[i,j]$ and diagonalizing it. Some of the results are shown in Fig.\  \ref{fig:timemodefunction2}, Fig.\  \ref{fig:BP}(b) and Fig.\  \ref{fig:BF}(b).

\begin{acknowledgments}
This work was partly supported by JST [Moonshot R\&D][Grant No.\ JPMJMS2064], JSPS KAKENHI (Grant No.\ 18H05207, No.\ 18H01149, and No.\ 20K15187), UTokyo Foundation, and donations from Nichia Corporation. M.E. acknowledges supports from Research Foundation for Opto-Science and Technology. T.S. acknowledges supports from The Forefront Physics and Mathematics Program to Drive Transformation (FoPM). The authors would like to thank Mr. Takahiro Mitani for careful proofreading of the manuscript.
\end{acknowledgments}

\bibliography{paper11.bib}

\end{document}